\documentclass[a4paper,twocolumn]{article}
\usepackage[utf8]{inputenc}
\usepackage{hyperref}
\usepackage{ytableau, array, amsmath}
\usepackage{float}
\usepackage{authblk}
\usepackage{graphicx}
\usepackage{lipsum}
\usepackage{blindtext}
\usepackage[caption=false]{subfig}
\graphicspath{ {./images/} }
\usepackage[margin=.5in]{geometry}
\usepackage{braket}
\begin{document}
\title{Tetraquark masses using the extension of Gursey- Radicati mass formula}
\author{\large Ankush Sharma
\thanks {ankushsharma2540.as@gmail.com}}
\author{\large Alka Upadhyay
\thanks {alka.iisc@gmail.com}}
\affil{School of Physics and Material Science, Thapar Institute of Engineering and Technology, Patiala-147004, India} 
\date{}
\twocolumn[
\begin{@twocolumnfalse}
\maketitle
\begin{abstract}
Recently, many exotic hadrons have been discovered at LHCb including the discovery of two singly heavy tetraquark structures $T_{c\bar{s}0}^a (2900)^{++}$ and $T_{c\bar{s}0}(2900)^0$ were observed in the $D_s^+\pi^+$ and $D_s^+\pi^-$ invariant mass spectra in two B-decay processes $B^+ \rightarrow D^-D_s^+\pi^+$ and $B^0 \rightarrow \bar{D^0}D_s^+ \pi^-$, respectively. To study the mass spectra of the tetraquark systems like $Q\bar{q}q\bar{q}$/$q\bar{q}q\bar{Q}$, $Q\bar{q}Q\bar{q}$/$qq\bar{Q}\bar{Q}$, $Q\bar{Q}q\bar{Q}$ and $QQ\bar{Q}\bar{Q}$, we have used the extension of the Gursey-Radicati mass formula for singly heavy to fully heavy tetraquark states. We have compared the mass spectra with the available experimental and theoretical data and studied their classification using the $SU(3)_f$ flavor, $SU(2)_s$ spin, $SU(6)_{sf}$ spin-flavor symmetry and the Young tableau technique. Also, decay channels and widths for the hidden-charm tetraquark states have been analyzed in the framework of the diquark-antidiquark model in terms of the coupling constant $F_{T{c\bar{c}}}^2$ (ranges from 0 to 1). We have predicted $J^P$ value for the newly predicted singly heavy tetraquarks by LHCb having quark content $u\bar{d}c\bar{s}$ and $\bar{u}dc\bar{s}$ and our analysis assumes that these states are s-wave states with $J^P$ value $0^+$ or $1^+$ and forming isospin triplet($I = 1$) and other assignments are also given to different possible states. 
 \end{abstract}
 \end{@twocolumnfalse}
 ]
\section{Introduction}
Recent discoveries of exotic hadrons by experimental facilities like LHCb, BaBar, and CDF have developed the theoretical interest in studying their dynamics using different theoretical aspects like effective theories, potential models, etc. Due to color confinement and asymptotic freedom, only color singlet states exist, such as mesons ($q\bar{q}$), baryons ($qqq$), tetraquarks ($q\bar{q}q\bar{q}$) mesons and pentaquarks ($qqqq\bar{q}$) baryons. Quantum chromodynamics also allows the existence of color singlet exotic hadrons such as compact multiquark states, hadronic molecules, and glueballs. Recently, in 2022, LHCb collaboration discovered two tetraquark structures having quark content $u\bar{d}c\bar{s}$ with a significance of 6.5 $\sigma$ and $\bar{u}dc\bar{s}$ with a significance of 8 $\sigma$ and a pentaquark structure $udsc\bar{c}$ with the significance of
15 $\sigma$, which is far beyond the 5 standard deviations required to claim the observation of a particle in particle physics. Two new tetraquark candidates $T_{c\bar{s}0}^a (2900)^{++}$ and $T_{c\bar{s}0}(2900)^0$ were observed in the $D_s^+\pi^+$ and $D_s^+\pi^-$ invariant mass spectra in two B-decay processes $B^+ \rightarrow D^-D_s^+\pi^+$ and $B^0 \rightarrow \bar{D^0}D_s^+ \pi^-$, respectively\cite{1}. The isospin and spin-parity quantum numbers are determined to be $(I)J^P = (1)0^+$. These two states should correspond to the two different charged states of the isospin triplet. The measured mass and width are $M_{exp} = 2908\pm 11 \pm 20$ MeV and $\Gamma_{exp}= 136 \pm 23 \pm 11 $ MeV. The $T_{c\bar{s}0}^a (2900)^{++}$ may be a flavor partner of the $0^+$ state $X^0(2900)$ (composed [$\bar{c}su\bar{d}$]) observed in the $D^-K^+$ final state in $B^+ \rightarrow D^+D^-K^+$ at LHCb in 2020. Also, in 2021, $Z_{cs}$(3985) and $Z_{cs}$(4003) tetraquarks having quark content $c\bar{c}s\bar{q}(c\bar{c}q\bar{s})$\cite{2,3} and $P_{cs}(4459)$ pentaquark state have been observed\cite{4}, where $q$ is the light quark $(u,d)$. The $Z_{cs}(3985)$ was fitted by the Breit-Wigner method, whose pole mass and width were determined to be $3982.5_{-2.6}^{+1.8}$ $\pm$ 2.1 MeV and $12.8_{-4.4}^{+5.3}$ $\pm$ 3.0 MeV, respectively; its $J^P$ quantum numbers are expected to be $1^+$\cite{2}. The $Z_{cs}(4003)$ have mass of $4003 \pm 6_{-14}^{+4}$ MeV and width of 131 $\pm$ 15 $\pm$ 26 MeV; its spin-parity ($J^P$) quantum numbers are expected to be $1^+$ with high significance of 15$\sigma$\cite{3}. The $P_{cs}$(4459) was discovered in the J/$\psi$ $\Lambda$ invariant mass distribution from an amplitude analysis of the $\Xi_b^-$ $\rightarrow$ $J/\psi$ $\Lambda$ $K^-$ decays. The observed structure is consistent with a hidden-charm pentaquark with strangeness, characterized by a mass of 4458.8 $\pm$ $2.9_{-1.1}^{+4.7}$ MeV and a width of 17.3 $\pm$ $6.5_{-5.7}^{+8.0}$ MeV. Its spin is expected to be 1/2 or 3/2, and its parity can be either -1 or +1\cite{4}.\\
In 2003, X(3872) was discovered by the Belle collaboration, which lies above the open charm threshold but had a narrow decay width of the order of 1.2 MeV via the following decay process,  $B^+ \rightarrow \psi + K^+$\cite{5}. They could also have the exotic contributions, referred to as charmonium-tetraquark states as  $B^+ \rightarrow X(3872) + K^+$\cite{6}.
\begin{figure}[H]
    \centering
    \includegraphics[width = 7cm]{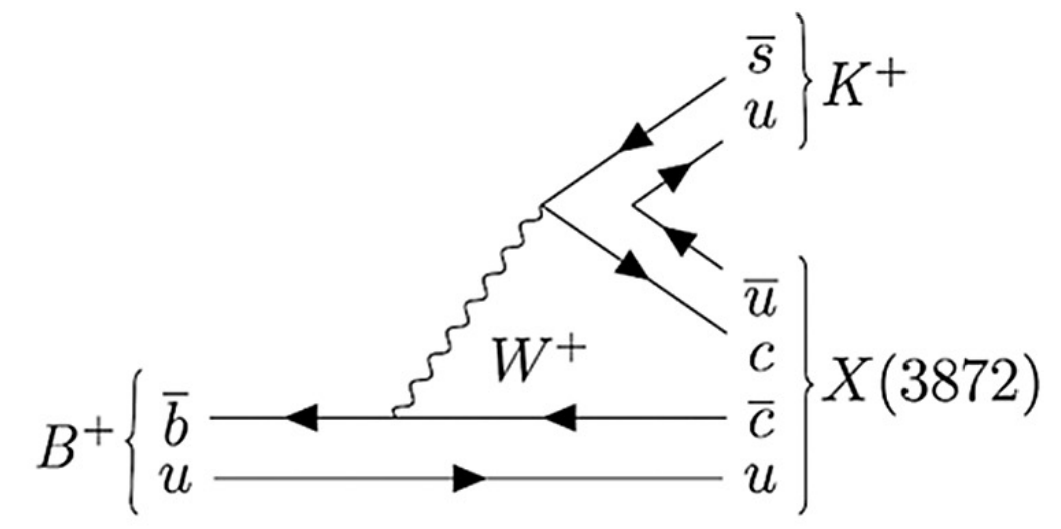}
    \caption{Production of X(3872) state\cite{65}}
    \label{fig:1}
\end{figure}
Two years later, BaBar collaboration discovered Y(4260) having spin-parity $1^{--}$ in the process $e^+e^-$ $\rightarrow$ $\gamma$ ($\pi^+$$\pi^-$J/$\psi$), has a width
of about 90 MeV, and it is seen to decay in $J/\psi \pi^+\pi^-$\cite{7}.  Also, in 2007, the Belle collaboration observed $Z^+(4430)$, a hidden charm tetraquark state through Y(4260) $\rightarrow \pi^- \pi^+ J/\psi$\cite{8}. In 2013, the BESIII Collaboration discovered $Z_c^+$(3900), which can decay into J/$\psi \pi^+$ through strong interactions\cite{9}. The LHCb collaboration reported more charmonium-like states with strange content in proton-proton collisions\cite{10}. The $Z_{cs}$(4000) is observed in the $B^+$ → J/$\psi \phi K^+$ decay, with mass and width 4003$\pm 6_{-14}^{+4}$ MeV and $131\pm 15 \pm 26$ MeV respectively, and the preferred spin-parity is $J^P = 1^+$. The X(4685), with $J^P = 1^+$ quantum numbers, decays to J/$\psi \phi$ final state with a high significance claimed by the collaboration.  Recently, the LHCb collaboration has reported four J/$\psi\phi$ structures,i.e X(4140), X(4274), X(4500) and X(4700) by studying the process $B^+\rightarrow$ J/$\psi\phi K^+$ using 3 $fb^{-1}$ data on $p\bar{p}$ collisions at $\sqrt{s}$ = 7 and 8 TeV having significance over 5 standard deviations\cite{11,12}. Their masses and decay widths have been determined as\cite{11}:
\begin{align*}
  M_{X(4140)} = (4146.5\pm 4.5_{-2.8}^{+4.6}) MeV
     \\
     \Gamma_{X(4140)} = (83 \pm 21_{-14}^{+21}) MeV
    \\
    M_{X(4274)} = (4273.3 \pm 8.3_{-3.6}^{+17.2}) MeV
    \\
    \Gamma_{X(4274)} = (56 \pm 11_{-11}^{+8}) MeV
    \\
    M_{X(4500)} = (4506 \pm 11_{-15}^{+12}) MeV,
\\
\Gamma_{X(4500)} = (92 \pm 21 _{-20}^{+21}) MeV
\\
M_{X(4700)} = (4704 \pm 10_{-24}^{+14}) MeV
\\
\Gamma_{X(4700)} = (120 \pm 31_{-33}^{+42}) MeV.
\end{align*}

The above results are obtained by Breit-Wigner parameterization of J/$\psi$$\phi$ mass fits. The first errors are statistical, and the second errors are systematic. These four J/$\psi$ structures have developed many theorists' attention after their production\cite{13, 14, 15, 16, 17, 18}. In 2020, the LHCb reported a narrow X(6900) structure in the J/$\psi$ pair invariant mass spectrum with a significance level of more than 5$\sigma$. The mass and width of the X(6900) resonance are measured to be\cite{19}:\\
\begin{center}
      $M = 6886 \pm 11 \pm 11$ MeV \\
    $\Gamma = 168 \pm 33 \pm 69$ MeV\\
\end{center}
  The charmonium-like states are also called XYZ states where X is the isospin-singlet state with $J^{PC} \neq 1^{--}$, Y is the isospin-singlet state with $J^{PC}= 1^{--}$, and Z is the isospin-triplet state\cite{20}. The masses of these states are above the open-charm thresholds. Due to the unexpected resonance parameters and decay channels, conventional quark models cannot describe these states. Therefore, they are good candidates for exotic states, such as hybrids, tetraquarks, molecules, etc.\cite{21,22,23}. Recently, BESIII Collaboration reported an enhancement for the $e^+e^- \rightarrow \gamma X(3872)$
production cross sections around 4.2 GeV\cite{24}, which tells us about the connection between X and Y states. Also, with the observation of a Y(4220) resonance in the process
$e^+e^-$ → $\pi^0 Z_c(3900)^0$\cite{25}. 
\\
 Many theoretical approaches are proposed, such as Born-Oppenheimer approximation\cite{26}, compact tetraquark model\cite{27}, Spectroscopy of pentaquark states\cite{28},  diquark-antidiquark model\cite{29}, compact pentaquark structures\cite{30}, hadro-charmonium model\cite{31}, heavy quark spin symmetry and chiral effective field theory are employed to study these tetraquarks and pentaquark states\cite{32}. In the past decade, many hidden charm and bottom tetraquark states have been studied by different models and theories for their masses, decay widths and branching ratio, etc.\cite{33,34,35,36,37,38,39,40,41,42,43,44,45,46,47,48,49}. T Mehen, in 2017, examined the excited doubly heavy baryons and tetraquarks using the implications of heavy quark-diquark symmetry and found the relationship between the masses of tetraquarks and baryon states by using the limits of HQET \cite{50}. Spectroscopy of fully heavy tetraquarks is carried out using the framework of NRQCD with OGE color Coulomb interaction \cite{51}, and masses of doubly heavy tetraquarks are calculated using the heavy quark effective theory \cite{52}.\\
 Many works have been proposed for pentaquarks using the Gursey-Radicati mass formula to calculate their mass spectrum. In ref.\cite{30},  by using the extension of Gursey-Radicati mass formula, hidden-charm pentaquark masses have been calculated by assigning them to an octet of SU(3) representation by using the extension of the Gursey-Radicati mass formula. In ref. \cite{53}, strange baryon spectrum and mass formula parameters have been calculated in ref. \cite{54}, where different fits have been employed to calculate the masses of pentaquark states. Thus G-R mass formula is a very useful tool for studying the mass spectra of exotic states. We have applied the extension of the Gursey-Radicati formula to evaluate the mass spectrum of tetraquark states by introducing some corrections into the parameters of the mass formula.\\
 In this work, the classification of tetraquarks is done using the SU(3) flavor and SU(2) spin and then using a large SU(6) spin-flavor representation. We have classified tetraquarks into octets based on their obtained masses. Several models have been proposed to classify tetraquarks, such as the diquark-antidiquark model, meson-meson molecule, etc. In section 3, using the extension of the Gursey-Radicati mass formula, the masses of tetraquark states have been calculated for each possible configuration. We analyzed and concluded at the end.
\section{The Classification of Tetraquark states}
We use the symmetry principle and the Young tableau technique to classify the tetraquark states. One can think of a tetraquark as a meson-meson molecule or consisting of two quarks and two anti-quarks as a diquark-antidiquark structure. Therefore, using the symmetry principle, the wavefunction of a tetraquark must be color singlet because it is composed of two pairs of fermions. Thus their total wave function must be anti-symmetric for the exchange of two quarks or two anti-quarks. In the case of flavor $SU(3)_f$ representation for the $q\bar{q}q\bar{q}$ system, by assigning a fundamental [3] to the quark and [$\bar{3}$] to the anti-quark, we have obtained following irreducible representations:
\begin{eqnarray}
    [3] \otimes [\bar{3}] \otimes [3] \otimes [\bar{3}] = [1] \oplus [8] \oplus [1] \oplus [8] \oplus [27] \\ \nonumber
    \oplus [8] \oplus [8] \oplus [10] \oplus [\overline{10}]   
\end{eqnarray}
It provides two singlets, four octets, a 27-plet and a 10 and $\bar{10}$ plets. For the case of fully-light tetraquarks, their classification into 10 and 27-plet is done into reference \cite{55}, and hidden-charm tetraquarks are classified into octets into reference \cite{29}. I = $0, \frac{1}{2}, \frac{3}{2}, 1, 2$ are the allowed isospin values and Y =$ 0, \pm 1, \pm 2$ are the allowed hypercharge values, where I = $\frac{3}{2}$, 2 and Y = $\pm$ 2 are exotic and do not occur for conventional $q\bar{q}$ mesons. As shown in equation (1) for the flavor, a similar classification is valid for the color wavefunction, and two possible color singlet representations for tetraquarks make the color a non-trivial quantum number for the tetraquarks. These two color singlet basis for tetraquarks i.e $\ket{6\bar{6}}_c$ and $\ket{\bar{3}3}_c$, their wave functions is given by\cite{57}: 
\begin{eqnarray}
\ket{6\bar{6}}_c = \frac{1}{2\sqrt{6}}\Large[(rb+br)(\bar{b}\bar{r} +\bar{r}\bar{b}) + (gr + rg)(\bar{g}\bar{r} + \bar{r}\bar{g}) \\ \nonumber+ (gb + bg)(\bar{b}\bar{g} + \bar{g}\bar{b})+ 2(rr)(\bar{r}\bar{r}) \\ \nonumber + 2(gg)(\bar{g}\bar{g}) + 2(bb)(\bar{b}\bar{b})\Large]
\end{eqnarray}
\begin{eqnarray}
\ket{\bar{3}3}_c = \frac{1}{2\sqrt{3}}\Large[(br - rb)(\bar{b}\bar{r} - \bar{r}\bar{b}) - (rg - gr)(\bar{g}\bar{r} - \bar{r}\bar{g}) \\ \nonumber + (bg - gb)(\bar{b}\bar{g} - \bar{g}\bar{b})\Large]
\end{eqnarray}
For $SU(2)_s$ spin symmetry, each quark is assigned by a fundamental [2] for both the quark as well as anti-quark and thus we have the following spin multiplets: 
\begin{eqnarray}  
[2] \otimes [2] \otimes [2] \otimes [2] = [1] \oplus [3] \oplus [1] \oplus [3] \oplus [3] \oplus [5]  
\end{eqnarray}
In the spin space, there are six spin bases, which are denoted by $\Xi_{SS_{z}}^{S_{12}S_{34}}$. Where $S_{12}$ stands for the spin quantum number for the diquark $(q_1q_2)$ (or antidiquark $(\bar{q_1q_2} )$), while $S_{34}$ stands for the spin quantum number for the antidiquark $(\bar{q_3}\bar{q_4})$
(or diquark $(q_3q_4)$). S is the total spin quantum number of the tetraquark $q\bar{q}q\bar{q}$ system, while $S_z$ stands for the third component of the total spin S. The spin wave functions $\Xi_{SS_{z}}^{S_{12}S_{34}}$ 
with a determined $S_z$ can be explicitly expressed as follows \cite{57}:
\begin{align*}
\Xi_{00}^{00} = \frac{1}{2}(\uparrow\downarrow\uparrow\downarrow - \uparrow\downarrow\downarrow\uparrow - \downarrow\uparrow\uparrow\downarrow + \downarrow\uparrow\downarrow\uparrow),\\
\Xi_{00}^{11} = \sqrt{\frac{1}{12}}(2\uparrow\uparrow\downarrow\downarrow - \uparrow\downarrow\uparrow\downarrow - \uparrow\downarrow\downarrow\uparrow - \downarrow\uparrow\uparrow\downarrow 
- \downarrow\uparrow\downarrow\uparrow 
 \\ + 2\downarrow\downarrow\uparrow\uparrow),\\
\Xi_{11}^{01} = \sqrt{\frac{1}{2}} (\uparrow\downarrow\uparrow\uparrow - \downarrow\uparrow\uparrow\uparrow), \\
\Xi_{11}^{10} = (\uparrow\uparrow\uparrow\downarrow - \uparrow\uparrow\downarrow\uparrow),\\
\Xi_{11}^{11} =\frac{1}{2}(\uparrow\uparrow\uparrow\downarrow + \uparrow\uparrow\downarrow\uparrow - \uparrow\downarrow\uparrow\uparrow - \downarrow\uparrow\uparrow\uparrow),\\
\Xi_{22}^{11} = \uparrow\uparrow\uparrow\uparrow.
\end{align*}
In the case of $SU(6)_{sf}$ spin-flavor representation of tetraquark system \cite{56}, it is represented as;
\begin{eqnarray}
    [6] \otimes [\bar{6}] \otimes [6] \otimes  [\bar{6}] = [1] \oplus [35] \oplus [405] \oplus [1] \oplus [35] \\ \nonumber \oplus [189] \oplus [35] \oplus [280] \oplus [35] \oplus [\overline{280}]
\end{eqnarray}
In $SU(6)_{sf}$ spin-flavor symmetry, the above representation is shown in subsection 2.1 by using the Young tableau method where fundamental [1] representation is used for quark, and $[1^{n-1}]$ is used for anti-quark. Now, to understand the spatial part of the wave function, we make use of the relative coordinates. Tetraquarks are made up of two quarks and two anti-quarks. Therefore we have to define three relative coordinates, as shown below.
\begin{align*}
  \Vec{r_{13}} &= \Vec{r_3} - \Vec{r_1} \\
  \Vec{r_{24}} &= \Vec{r_4} - \Vec{r_2} \\
  \Vec{r_{12-34}} &= \Vec{r_{CM24}} - \Vec{r_{CM13}} \\&= \frac{m_2 \Vec{r_2}+ m_4\Vec{r_4}}{m_2 + m_4} - \frac{m_1\Vec{r_1} +m_3\Vec{r_3}}{m_1+m_3}
\end{align*}
Which is only a possible choice of coordinates. In the tetraquark case, we have four different spins and three different orbital angular momentum values, and thus total angular momentum J can be obtained by combining the spins and orbital angular momenta as shown in figure 2 \cite{58}\\
  \begin{figure}[ht]
      \centering
    \includegraphics[width = 0.8\linewidth]{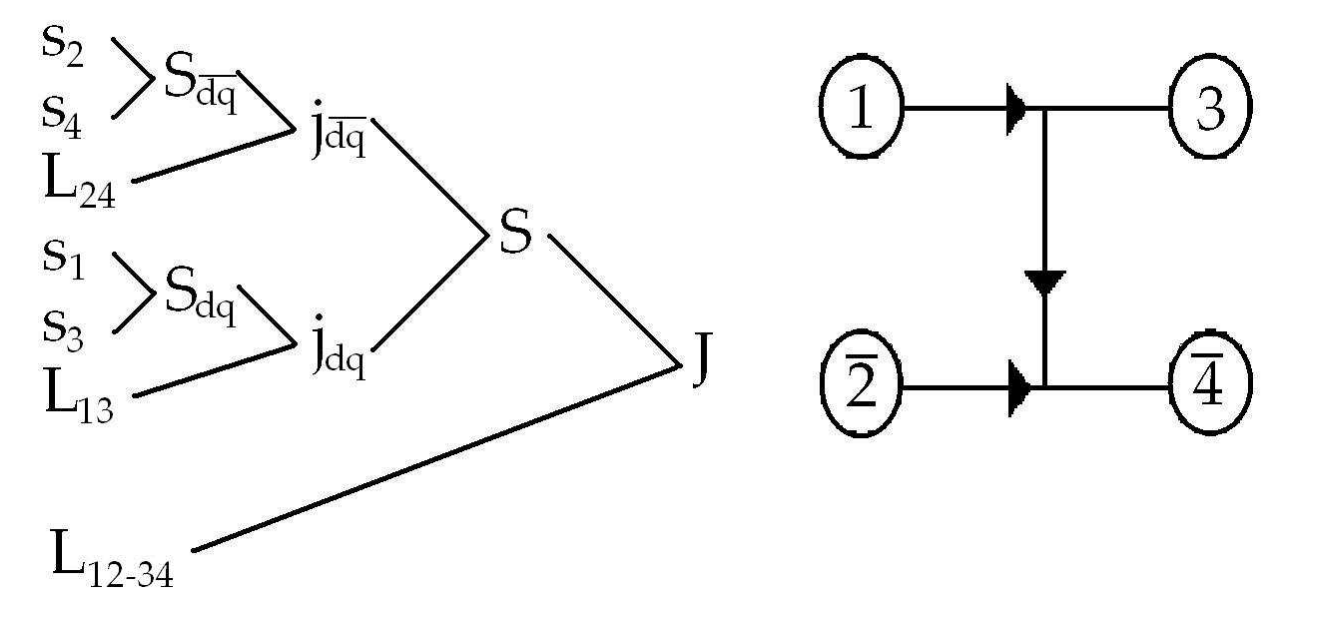}
      \caption{Angular momentum and Parity quantum number\cite{58}}
      \label{fig:2}
  \end{figure}
Parity for a tetraquark system is the product of the intrinsic parities of the quarks and the antiquarks times the factors coming from the spherical harmonics\cite{59}.
\begin{eqnarray}
P &= P_q P_q P_{\bar{q}} P_{\bar{q}} (-1)^{L_{13}} (-1)^{L_{24}} (-1)^{L_{12-34}} \nonumber \\ &= (-1)^{L_{13} + L_{24} +L_{12-34}}
\end{eqnarray}
Therefore we can say that parity for the ground state, i.e., S-wave (l=0), is positive or natural. Let's discuss the possible permutations of color, spin-flavor, and spatial parts of tetraquarks. As only those physical states exist, which are color singlets. In the case of tetraquarks, there are two color singlets that can be only antisymmetric (A) or symmetric (S) by considering the permutation $S_2$ symmetry\cite{59}.
\begin{center}
     $(qq)$ in $[\bar{3}]_C$ (A) and $(\bar{q}\bar{q})$ in $[3]_C$ (A)\\
     $(qq)$ in $[6]_C$ (S) and $(\bar{q}\bar{q})$ in $[\bar{6}]_C$ (S) \\
\end{center}
Now, we study the permutation symmetry of the spatial part of the tetraquarks by taking a couple of quarks and anti-quarks.
\begin{center}
    $(qq)$ with $L_{13}$ even (S), $(\bar{q}\bar{q})$ with $L_{24}$ even (S)\\
    $(qq)$ with $L_{13}$ odd (A), $(\bar{q}\bar{q})$ with $L_{24}$ odd (A)\\
     $(qq)$ with $L_{13}$ even (S), $(\bar{q}\bar{q})$ with $L_{24}$ odd (A)\\
      $(qq)$ with $L_{13}$ odd (A), $(\bar{q}\bar{q})$ with $L_{24}$ even (S)
\end{center}
Permutation symmetry of the spatial part derives from the parity of a couple of quarks and antiquarks which is given as: $P_{qq}= P_q P_q (-1)^{L_{13}} = (-1)^{L_{13}}$ and $P_{\bar{q}\bar{q}}= P_{\bar{q}} P_{\bar{q}} (-1)^{L_{24}} = (-1)^{L_{24}}$.\\
The permutation symmetry of the $SU(6)_{sf}$ representations for a couple of quarks is written below.
\begin{center}
 $[15]_{sf}$(A), which means symmetric spin $(S_{dq} = 1)$ and antisymmetric flavor ($[\bar{3}]_f$) or antisymmetric spin $(S_{dq}= 0)$ and symmetric flavor $([6]_f)$. \\
$[21]_{sf}$ (S), which means symmetric spin $(S_{dq} = 1)$ and symmetric flavor $([6]_f)$ or antisymmetric spin
$(S_{dq} = 0)$ and antisymmetric flavor $([\bar{3}]_f)$. 
\end{center}
The flavor wave function of tetraquarks for all the multiplets is written in reference \cite{60}. Hidden charm tetraquark states are classified by the flavor $SU(3)_f$ symmetry under which three light quarks form a triplet 3 representation and charm quarks form a singlet \cite{50,51,52}. The flavor components of tetraquarks in flavor SU(3) symmetry are given below\cite{29}, and their octet is shown in figure 3:\\
\begin{gather}
 Z_c^0 = \frac{1}{\sqrt{2}}(u\bar{u}-d\bar{d})c\bar{c}, X = \frac{1}{\sqrt{6}}(u\bar{u}+d\bar{d}-2s\bar{s})c\bar{c} \nonumber \\
Z_c^+ = u\bar{d}c\bar{c},  Z_c^- = d\bar{u}c\bar{c}, Z_{cs}^+ = u\bar{s}c\bar{c} \nonumber \\
Z_{cs}^- = s\bar{u}c\bar{c},Z_{cs}^0 = d\bar{s}c\bar{c}, \bar{Z}_{cs}^0 = s\bar{d}c\bar{c} \nonumber \\ 
X^{'} = \frac{1}{\sqrt{3}}(u\bar{u}+d\bar{d}+s\bar{s})c\bar{c} \nonumber
\end{gather}
\begin{figure}
    \centering
    \includegraphics[width =9cm]{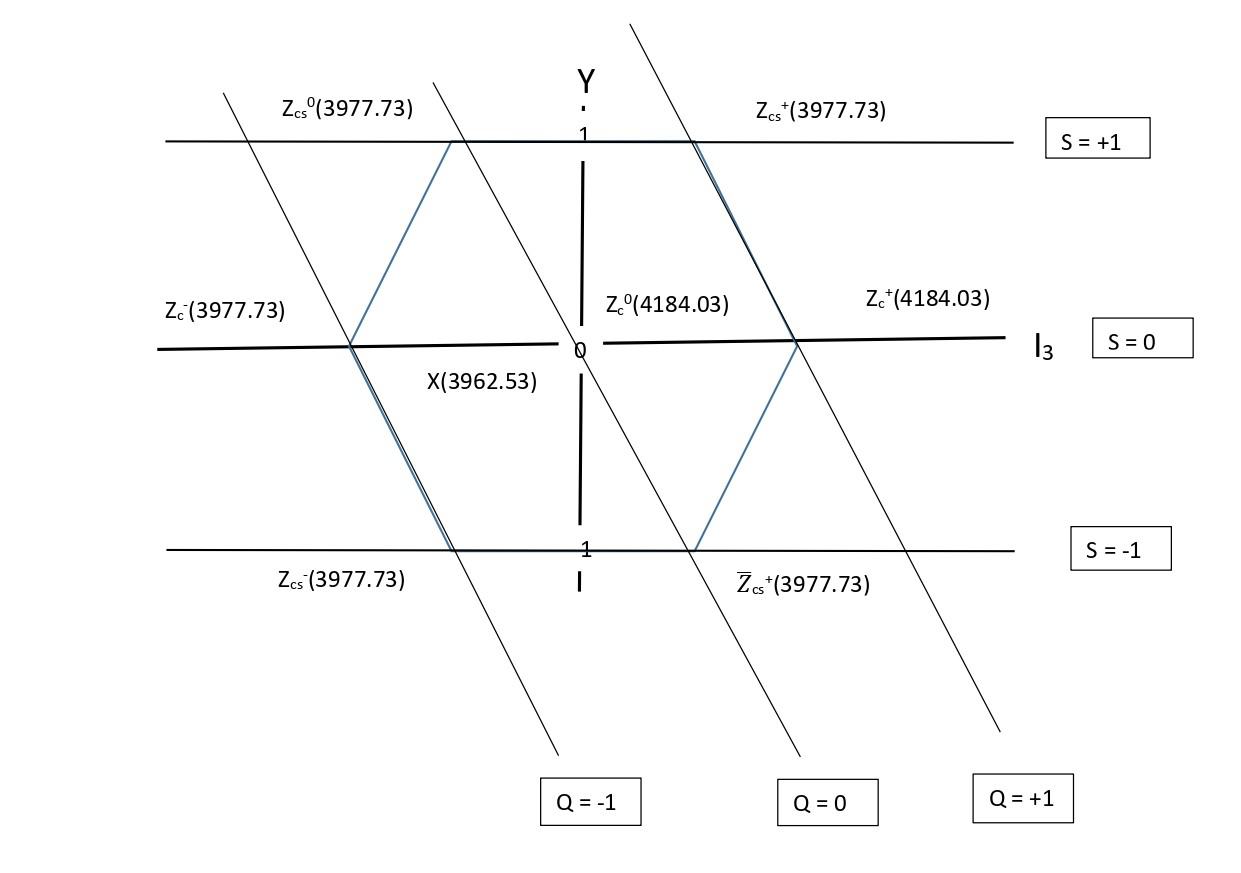}
    \caption{Octet of the charmonium tetraquark states: each state is labeled as $Z^i(M)$, where i is the tetraquark's electric charge and M is the predicted mass.}
    \label{fig:3}
\end{figure}
Also, in case of 5 flavors ($u$, $d$, $s$, $c$, $b$), the hypercharge is defined as: \\
\begin{eqnarray}
   Y = B + S + C + B' + T' 
\end{eqnarray}
where $B$ is the barionic number, $S$ is the strangeness, $C$ is the charmness, $B'$ is the bottomness and $T'$ are the topness quantum numbers for quarks and antiquarks.
By making use of the Young tableau technique to construct the allowed $SU_{sf}$(6) representations for the tetraquark system, denoting with a box the fundamental representation of SU(n), with n = 2,3,6 for the spin, flavor (or color), and spin-flavor degrees of freedom, respectively. The quark transforms as the fundamental representation [1] under SU(n), whereas the antiquark transforms as the conjugate representation [$1^{n-1}$] under SU(n). Also, as their corresponding representation. e.n SU(3) and SU(2) are also shown by considering the inner product of these two representations as SU(6) representations. The spin-flavor classification for the quark and antiquark are given by:
\\
\begin{table}[H]
    \centering
    \begin{tabular}{ccccc}
   $SU_{sf}(6)$ & $\supset$ & $SU_f(3)$ & $\otimes$ & $SU_s(2)$\\
   & & & & \\
quark [\ 1 ]\ & $\supset$ \ & [\ 1 ]\ & $\otimes$ & [\ 1 ]\ \\ 
   & & & & \\
   \begin{ytableau}
     \none &
\end{ytableau} & $\supset$ &  \begin{ytableau}
     \none &
\end{ytableau} & $\otimes$
& \begin{ytableau}
     \none &
\end{ytableau} \\
 & & & & \\
 antiquark [\ 11111 ]\  & $\supset$ &  [\ 11 ]\  & $\otimes$ & [\ 1 ]\ \\
 & & & & \\
   \begin{ytableau}
     \none  &  \\
     \none  &  \\
     \none  &  \\
     \none  &  \\          
     \none  &  \\
\end{ytableau} & $\supset$ &
 \begin{ytableau}
     \none & \\
     \none & \\
  \end{ytableau} & $\otimes$ 
 & \begin{ytableau}
     \none & 
   \end{ytableau}
    \label{tab:my_label}
\end{tabular}
\end{table}
We used the inner product of single quark states on the right-hand side. The spin-flavor states of multi-quark systems can be obtained by taking the outer product of the representations of the quarks and anti-quarks.
\subsection{$q\bar{q}q\bar{q}$ System}
In this section, we have make use of the higher $SU(6)_{sf}$ spin-flavor symmetry to study the classification of tetraquark system into multiplets which are not possible into SU(3) representation. Therefore by using the Young table technique for the tetraquark system\cite{54}, we have:\\
\begin{align*}
\begin{ytableau}
\none & \\
\none & \\
\none & \\
\none & \\
\none & \\
\end{ytableau} \otimes
 \begin{ytableau}
\none & \\
\none & \\
\none & \\
\none & \\
\none & \\
\end{ytableau} 
\otimes \begin{ytableau}  \none & \\
\end{ytableau} \otimes \begin{ytableau}  \none & \\
\end{ytableau}  =
\begin{ytableau}
\none & & & & \\
\none & & \\
\none & & \\
\none & & \\
\none & & \\
\end{ytableau} \\ \oplus  \begin{ytableau}
\none & & \\
\none & & \\
\none & & \\
\none & & \\
\none & & \\
\none & & \\
\end{ytableau}  \oplus \begin{ytableau}
\none & & & \\
\none & & & \\
\none & & \\
\none & & \\
\none & & \\
\end{ytableau} \oplus 
\begin{ytableau}
\none & & & & \\
\none & & \\
\none & & \\
\none & & \\
\none & \\
\none & \\
\end{ytableau} \oplus \\
\begin{ytableau}
\none & & & \\
\none & & & \\
\none & & \\
\none & & \\
\none & \\
\none & \\
\end{ytableau} \oplus
4\begin{ytableau}
\none & & & \\
\none & & \\
\none & & \\
\none & & \\
\none & & \\
\none & \\
\end{ytableau} \oplus \begin{ytableau}
\none & & \\
\none & & \\
\none & & \\
\none & & \\
\none & & \\
\none & & \\
\end{ytableau} 
\end{align*}
Which can be represented as : 
\begin{eqnarray}
[11111]_{\bar{6}} \otimes [11111]_{\bar{6}} \otimes [1]_6 \otimes [1]_6 = \\ \nonumber  [42222]_{405} \oplus [33222]_{280} \oplus [422211]_{\overline{280}} \\ \nonumber \oplus [332211]_{189}  \oplus 4 [322221]_{35}  \oplus 2 [222222]_{1}  
\end{eqnarray}
The complete classification of the tetraquarks involves the analysis of the flavor and spin content of each representation, i.e., decomposition of $SU(6)_{sf}$ into those of $SU(3)_f$ $\otimes$ $SU(2)_s$ is shown below as \cite{61}:\\
\begin{gather*} 
[189] = [8,5] \oplus [10,3] \oplus [27,1] \oplus [\bar{10},3] \oplus 2[8,3] \\ \oplus [8,1] \oplus [1,1] \oplus [1,5].\\
    [280] = [10,5] \oplus [8,5] \oplus [27,3] \oplus [10,3] \oplus 2[8,3] \\ \oplus [10,1] \oplus [10,1]\oplus [8,1] \oplus [1,3]. \\
 [\bar{280}] = [10,5] \oplus [8,5] \oplus [27,3] \oplus [10,3] \oplus 2[8,3] \\ \oplus [10,1] \oplus [10, 1] \oplus [8, 1] \oplus [1, 3].\\
[405] = [1,1] \oplus [1,5] \oplus [8,5] \oplus 2[8,3] \oplus [27,1] \oplus [8,1] \\ 
\oplus [27,3] \oplus [10,3] \oplus [10,3] \oplus [27,5].\\
[1] = [1,1]. \\
[35] = [1,3] \oplus [8,1] \oplus [8,3].
\end{gather*}
Therefore in $SU(6)_{sf}$ representation, each multiplet is reducible in terms of multiplets of $SU(3)_f$ and $SU(2)_s$ and their tensor sum, which consists of singlets, octets, decouplets, and 27-plets of flavor SU(3) and singlets, triplets and pentets of spin SU(2) in groups gives us the exact size of spin-flavor SU(6) multiplets.
\section{The extension of the Gursey-Radicati mass formula}
We used an extension of the Gursey-Radicati mass formula to find the mass splitting between different multiplets\cite{30}. Generalization of this mass formula in terms of $SU(6)_{sf}$ is made using the Casimir operator for different quantum numbers. Assuming that the GR formula's coefficients are identical and have the same value for different quark systems, the authenticity of this mass formula is checked on three quark systems. The original mass formula in terms of the Casimir operator is given by :
\begin{eqnarray}
    M_{GR} = M_0 + C C_2 [SU_s(2)] + D C_1[U_Y(1)] \\ \nonumber + E [C_2[SU_I(2)]- \frac{1}{4}(C_1[U_Y(1)])^2]
\end{eqnarray}
Where, $C_2[SU_S(2)]$ and $C_2[SU_I(2)]$ are the SU(2) Casimir operators for spin and isospin, respectively, and $C_1[U_Y(1)]$ is the Casimir for the U(1) subgroup generated by the hypercharge Y. In the framework of the CQM, the underlying symmetry is provided by SU(6), therefore the most general formula that can be written based on a broken SU(6) symmetry: 
\begin{eqnarray}
   M = M_0 + AC_2[SU_{SF}(6)] + B C_2[SU_F(3)] \\ \nonumber + CC_2[SU_S(2)] + D C_1[U_Y(1)] + \\ \nonumber E(C_2[SU_I(2)] - \frac{1}{4}(C_1[U_Y(1)])^2) 
\end{eqnarray}
By putting the eigenvalues of the Casimir operator into the mass formula,
   $\langle{C_2[SU_I(2)]}\rangle = I(I+1)$,  $\langle{C_1[U_Y(1)]}\rangle = Y$,  $\langle{C_2[SU_S(2)]}\rangle = S(S+1)$,\\
Thus the simplest form of the GR formula, which differentiates the different multiples of $SU_{f}(3)$, is
\begin{eqnarray}
    M_{GR} = M_0 + A s(s+1) + DY+ G C_2(SU(3)) \\ \nonumber + E[I(I+1) - \frac{1}{4} Y^2]
\end{eqnarray}
The above form of GR mass formula is initially given to calculate the baryon masses, which is extended to exotic states like tetraquarks and pentaquarks shown below :
\begin{eqnarray}
    M_{GR} = \xi M_0 + AS(S+1) +DY + E[I(I+1)\\ \nonumber -\frac{1}{4}Y^2]  + GC_2(SU(3)) +F N_i
\end{eqnarray}
 Equation (12) is known as the extension of the Gursey-Radicati mass formula. Where $M_0$ is the scale parameter, each quark contributes $\frac{1}{3}M_0$ to the whole mass, and $\xi$ is the correction factor to the value of $M_0$. Its value is based on the number of quarks involved in the system. for baryons, its value is 1; for tetraquarks, its value is 4/3; and for pentaquark states, its value is 5/3. S, I, and Y are the spin, isospin, and hypercharge, respectively. At the same time, $C_2(SU(3)$ is the eigenvalue of the $SU_f$ Casimir operator. $N_i$(i stands for c or b quark) accounts for the counter of c quark or b quark or $\bar{c}$ antiquark or $\bar{b}$ antiquark, basically tells us about the mass difference between charm(anti-charm) quark and the light quarks.\\
We use information about well-known baryon states to find the coefficients A, D, E, G, and F and the scale parameter $M_0$. The mass difference between the two of them gives us the relations between the different coefficients, and fitting them all into the mass formula produces the best-fit mass spectrum for all the ground state baryons. Table 1 shows their quantum number assignments, and table 2 shows the values of parameters in the GR mass formula, along with their associated uncertainties. 
\begin{table*}[ht]
    \centering
    \begin{tabular}{|c|c|c|c|c|c|c|}
    \hline
       Baryons & $SU_f(3)$ & $C_2(SU(3)$ & S & Y & I & $N_c$ \\
       \hline
       N(940) & $[21]_8$ & 3 & $\frac{1}{2}$ & 1 & $\frac{1}{2}$ & 0 \\
$\Lambda^0(1116)$ & $[21]_8$ & 3 & $\frac{1}{2}$ & 0 & 0 & 0 \\ 
$\Sigma^0(1193)$ & $[21]_8$ & 3 & $\frac{1}{2}$ & 0 & 1 & 0 \\
$\Xi^0(1315)$ & $[21]_8$ & 3 & $\frac{1}{2}$ & -1 & $\frac{1}{2}$ & 0 \\
$\Delta^0(1232)$ & $[3]_{10}$ & 6 & $\frac{3}{2}$ & 1 & $\frac{3}{2}$ & 0 \\
$\Sigma^0(1385)$ & $[3]_{10}$ & 6 & $\frac{3}{2}$ & 0 & 1 & 0 \\
$\Xi^0(1530)$ & $[3]_{10}$ & 6 & $\frac{3}{2}$ & -1 & $\frac{3}{2}$ & 0 \\
$\Omega^{-}(1672)$ & $[3]_{10}$ & 6 & $\frac{3}{2}$ & -2 & 0 & 0 \\ 
$\Lambda_c^+$(2286) & $[11]_3$ & $\frac{4}{3}$ & $\frac{1}{2}$ & $\frac{2}{3}$ & 0 & 1 \\ 
$\Sigma_c^0$(2455) & $[2]_6$ & $\frac{10}{3}$ & $\frac{1}{2}$ & $\frac{2}{3}$ & 1 & 1 \\
$\Xi_c^0$(2471) & $[11]_3$ & $\frac{4}{3}$ & $\frac{1}{2}$ & $-\frac{1}{3}$ & $\frac{1}{2}$ & 1 \\ 
$\Xi_c^{'0}$(2576) & $[2]_6$ & $\frac{10}{3}$ & $\frac{1}{2}$ & $-\frac{1}{3}$ & $\frac{1}{2}$ & 1 \\ $\Omega_c^0$(2695) & $[2]_6$ 
& $\frac{10}{3}$ & $\frac{1}{2}$ & $-\frac{4}{3}$ & 0 & 1 \\
$\Omega_c^{*0}$(2770) & $[2]_6$ & $\frac{10}{3}$ & $\frac{3}{2}$ & $-\frac{4}{3}$ & 0 & 1 \\
$\Sigma_0^{*0}$(2520) & $[2]_6$ & $\frac{10}{3}$ & $\frac{3}{2}$ & $\frac{2}{3}$ & 1 & 1 \\
$\Xi_c^{*0}$(2645) & $[2]_6$ & $\frac{10}{3}$ & $\frac{3}{2}$ & $-\frac{1}{3}$ & $\frac{1}{2}$ & 1 \\
\hline
    \end{tabular}
    \caption{Quantum number assigned for the baryons written in table 3.}
   \end{table*} 

   \begin{table*}[ht]
       \centering
       \begin{tabular}{|c|c|c|c|c|c|c|}
       \hline
           & $M_0$ & A & D & E & F & G  \\
           \hline
        Values[MeV] & 940.0 & 23.0 & -158.3 & 32.0 & 1354.6 & 52.5 \\
        Uncertainties [MeV] & 1.5 & 1.2 & 1.3 & 1.3 & 18.2 & 1.3 \\
        \hline
       \end{tabular}
       \caption{Values of parameters in the GR mass formula with corresponding uncertainties}
   \end{table*}
   
Different fits are also employed for the parameters used in the mass formula in reference\cite{54}. The distinct eigenvalues of the Casimir operator $C_2(SU(3))$ cause mass splitting between the different multiplets of $SU_f(3)$,  and mass splitting is proportional to the coefficient G of the mass formula. Now we apply this mass formula to predict the masses of tetraquarks, which also helps us to predict the spin states of numerous tetraquarks with unknown $J^P$ values. 
\section{Decay Width}
We will now explore the possible decay channels of the hidden-charm tetraquark states, which helps us to observe the other predicted state of the octet (figure 3). In the framework of the diquark-antidiquark model, decay width evaluation for hidden-charm tetraquarks is carried out by considering their decay to D-mesons. As hidden-charm tetraquarks lie above the $D\bar{D}$ threshold, they can decay into two charmed mesons. Fields for D mesons have been assigned as $P_D$, $P_{\bar{D}}$ for $0^-$ and $P_{D^*}$ and $P_{\Bar{D^*}}$ for $1^-$ mesons respectively. We have studied the strong decays of a few hidden-charm tetraquarks to D mesons using the formulae\cite{29}. We have calculated its decay width by considering the possible decay channel for the tetraquark state. Decays of four different hidden-charm states denoted as $T_{cc}^+$, $T_{cc}^0$, $T_{cc}^{+'}$ and $T_{cc}^{0'}$ having quark content $u\bar{d}c\bar{c}$, $u\bar{u}c\bar{c}$, $u\bar{s}c\bar{c}$ and $s\bar{d}c\bar{c}$ respectively, have been calculated listed along with their quantum numbers and decay channels in table 3. \\ 
For tetraquarks having positive parity, we have transition amplitudes for two-body decays as:
 \begin{align*}
\mathcal{M}(T_{c\bar{c}}[0^+]\rightarrow D\bar{D}) = m_{T_{c\bar{c}}} F_{T_{c\bar{c}}[0^+]D\bar{D}}, \\
\\
\mathcal{M}(T_{c\bar{c}}[0^+]\rightarrow D^* \bar{D}^*) = \epsilon_{D^*,\mu}\epsilon_{\bar{D}^*,\nu}\times \\(g^{\mu\nu} - P_{\bar{D}^*}^{\mu}P_{D^*}^{\nu}{P_{D^*} P_{\bar{D^*}}})  \times m_{T_{c\bar{c}}} F_{T_{c\bar{c}}[0^+]D^*\bar{D}^*},\\
\\
\mathcal{M}(T_{c\bar{c}}[1^+]\rightarrow D\bar{D}) = \epsilon_{T_{c\bar{c}},\mu} \epsilon_{\bar{D}^*,\nu} (g^{\mu\nu} - \frac{P_{\bar{D}^*}^{\mu}P_{T_{c\bar{c}}}^{\nu}}{P_{T_{c\bar{c}}} P_{\bar{D^*}}}) \\ \times \frac{1}{\sqrt{3}}m_{T_{c\bar{c}}} F_{T_{c\bar{c}}[1^+]D\bar{D}^{*}},\\
\\
\mathcal{M}(T_{c\bar{c}}[2^+]\rightarrow D^{*}\bar{D}^{*}) = (- \frac{P_{\bar{D}^{*}}^{\beta} P_{D^{*}}^{\nu} \epsilon_{T_{c\bar{c}}}^{\mu\beta}}{P_{D^*} P_{\bar{D}^*}} - \frac{P_{\bar{D}^*}^{\mu} P_{D^*}^{\alpha} \epsilon_{T_{c\bar{c}}}^{\alpha\nu}}{P_{D^*} P_{\bar{D}^*}} + \\ \frac{P_{\bar{D}^*}^\mu P_{D^*}^\nu \epsilon_{T_{c\bar{c}}}^{\alpha\beta} P_{\bar{D}^*}^\beta P_{D^*}^\alpha}{(P_{D^*}.P_{\bar{D}^*})^2} +\epsilon_{T_{c\bar{c}}}^{\mu\nu} ) \epsilon_{D^*,\mu} \epsilon_{\bar{D}^*,\nu}\frac{m_{T_{c\bar{c}}} F_{T_{c\bar{c}}[2^+]D\bar{D}^*}}{\sqrt{5}} 
\end{align*}
  Where $F_{T_{c\bar{c}}D(D^*)\bar{D}(\bar{D}^*)}$ denotes the effective coupling to the tetraquark is the same as used in the reference \cite{62}. For the tetraquarks with $J^P = 1^{--}$, we have the two body decay amplitudes:
  \begin{align*}  
  \mathcal{M}(T_{c\bar{c}}[1^{--}]\rightarrow D\bar{D}) =  \epsilon_{T_{c\bar{c}}}
   (P_D - P_{\bar{D}}) \frac{F_{T_{c\bar{c}}[1^{--}]D\bar{D}}}{\sqrt{3}}\\
   \\
   \mathcal{M}(T_{c\bar{c}}[1^{--}]\rightarrow D\bar{D}^*) = 
 \epsilon_{T_{c\bar{c}}}^\mu \epsilon_{\bar{D}^*}^\nu P_{D}^\rho P_{\bar{D}}^\sigma \epsilon_{\mu \nu \rho \sigma} \times \\
   \frac{F_{T_{c\bar{c}}[1^{--}]D\bar{D}^*}}{\sqrt{3}m_{T_{c\bar{c}}}}\\
   \\
    \mathcal{M}(T_{c\bar{c}}[1^{--}]\rightarrow D^*\bar{D}^*) =  \epsilon_{T_{c\bar{c}}, \mu} \epsilon_{\bar{D}^*,\rho} \epsilon_{D^*,\nu} \frac{F_{T_{c\bar{c}}[1^{--}]D^*\bar{D}^*}}{\sqrt{3}} \\ \times (g^{\mu \rho}(- P_{T_{c\bar{c}}} - P_{\bar{D}^*})^\nu + g^{\mu\nu} P_{T_{c\bar{c}}}^\rho + g^{\mu\nu}P_{D^*}^\rho \\
    + g^{\rho\nu}(P_{\bar{D}^*} - P_{D^*})^\mu)
 \end{align*}
\\
The decay width of $T_{c\bar{c}} \rightarrow D(D^*) + \bar{D}(\bar{D}^*)$ can be written as \\
\\
 $\Gamma(T_{c\bar{c}} \rightarrow D(D^*) + \bar{D}(\bar{D}^*)) = \frac{|p|}{8\pi m_{T_{c\bar{c}}}^2} |\mathcal{M}|^2$ \\
 \\
 Where \\
 \\
$ |p| = \frac{\sqrt{(m_{T_{c\bar{c}}}^2 - (m_1 -m_2)^2)(m_{T_{c\bar{c}}}^2 - (m_1 + m_2)^2)}}{2m_{T_{c\bar{c}}}} $\\
\\
$|p|$ is the momentum modulus of the final charmed meson in the tetraquark rest frame, and $m_1$ and $m_2$ are the final charmed mesons. The corresponding ratios for the tetraquarks with positive parity are
\begin{align*}
\frac{\Gamma(T_{c\bar{c}}[0^+] \rightarrow D\bar{D})}{F_{T_{c\bar{c}[0^+]D\bar{D}}}^2|p|} =& \frac{1}{8\pi} \\
\\
\frac{\Gamma(T_{c\bar{c}}[0^+] \rightarrow D^*\bar{D^*})}{F_{T_{c\bar{c}[0^+]D^*\bar{D^*}}}^2|p|} =& \frac{3m_{T_{c\bar{c}}}^4 + 8 m_{T_{c\bar{c}}}^2 |p|^2 + 48 |p|^4}{8\pi(m_{T_{c\bar{c}}}^2 + 4|p|^2)^2}\\
\\
\frac{\Gamma(T_{c\bar{c}}[1^+] \rightarrow D\bar{D^*})}{F_{T_{c\bar{c}[1^+]D\bar{D^*}}}^2|p|} =& \frac{h_1}{12\pi(m_{T_{c\bar{c}}}^2 + m_{D^*}^2- m_D^2)^2}\\
\\
\frac{\Gamma(T_{c\bar{c}}[2^+] \rightarrow D^*\bar{D^*})}{F_{T_{c\bar{c}[2^+]D^*\bar{D^*}}}^2|p|} =& \frac{m_{T_{c\bar{c}}}^8}{\pi(m_{T_{c\bar{c}}}^2 + 4|p|^2)^2}(\frac{1}{8} + \frac{|p|^2 h_2}{15 m_{T_{c\bar{c}}}^8})\\
 \end{align*}
    where 
    \begin{align*}
    h_1 =& m_{T_{c\bar{c}}}^4 + 4 m_{T_{c\bar{c}}}^2 m_{D^*}^2 +m_{D^*}^4  -2 m_D^2(m_{D^*}^2  + m_{T_{c\bar{c}}}^2) + m_D^4,\\
    h_2 =& 15 m_{T_{c\bar{c}}}^6 + 76 m_{T_{c\bar{c}}}^4 |p|^2 + 176 m_{T_{c\bar{c}}}^2 |p|^4 + 224 |p|^6
 \end{align*}
The similar ratios for the tetraquarks with $J^P = 1^{--}$ are \\
\begin{eqnarray*}
   \frac{\Gamma(T_{c\bar{c}}[1^{--}] \rightarrow D\bar{D})}{F_{T_{c\bar{c}[1^{--}]D\bar{D}}}^2|p|^3} = \frac{1}{6 \pi m_{T_{c\bar{c}}}^2} \\
   \frac{\Gamma(T_{c\bar{c}}[1^{--}] \rightarrow D\bar{D^*})}{F_{T_{c\bar{c}[1^{--}]D\bar{D^*}}}^2 {|p|}^3} = \frac{1}{12\pi m_{T_{c\bar{c}}}^2}\\
   \frac{\Gamma(T_{c\bar{c}}[1^{--}] \rightarrow D^*\bar{D^*})}{F_{T_{c\bar{c}[1^{--}]D^*\bar{D^*}}}^2|p|^3} = \frac{m_{T_{c\bar{c}}}^4 - \frac{104}{9}|p|^2 + \frac{48}{9}|p|^4}{2\pi m_{T_{c\bar{c}}}^2 (m_{T_{c\bar{c}}}^2 - 4|p|^2)^2}  
   \end{eqnarray*}
\begin{table*}[ht]
    \centering
    \begin{tabular}{|c|c|c|c|c|c|c|}
    \hline
     Quark content & Mass & Decay Channel & S & I & Y & Decay Width \\
     \hline
     $u\bar{d}c\bar{c}$ & 4120.03 & $\bar{D}^0 D^+$ & 0 & 0 & 0 & 34.63$F_{T_{c\bar{c}}}^2$ \\
     \hline
     $u\bar{u}c\bar{c}$ & 4166.03 & $D\bar{D}^*$ & 1 & 0 & 0 & 29.28$F_{T_{c\bar{c}}}^2$ \\
     \hline
     $u\bar{s}c\bar{c}$ & 3977.73 & $\bar{D^0}D_s^+$ & 0 & 1/2 & 1 & 29.02$F_{T_{c\bar{c}}}^2$ \\
     \hline
     $s\bar{d}c\bar{c}$ & 4023.73 & $D_s^- D^{+*}$ & 1 & 1/2 & 1 & 14.77$F_{T_{c\bar{c}}}^2$ \\
     \hline
     \end{tabular}
    \caption{Decay Channels and Decay Width of Tetraquark states}
    \label{tab: table 3 }
\end{table*}
\section{Analysis}
In this work, the classification of tetraquarks is carried out using the $SU(3)_f$ flavor, $SU(2)_s$ spin, and then using the large $SU(6)_{sf}$ spin flavor representation. We have defined all possible combinations for the tetraquark's flavor, spin, color, and space wave functions.  We have considered the octets of the $SU(3)_f$ flavor representation and evaluated the masses for singly heavy to fully heavy tetraquarks (having $N_c$ and $N_b$ ranging from 1 to 4) using the extension of the Gursey-Radicati mass formula. This mass formula is firstly applied to well-known baryon states, and all the parameters ($A$, $D$, $E$, $G$, and $F$) used in the mass formula are calculated using this information listed in table 2. These mass formula parameters are evaluated by analyzing the difference in masses of well-known baryons states. Then we assumed that these parameters are fundamental and applicable to all the hadronic states. By considering all possible combinations of the quantum numbers and data feeding of these parameters ($M_0$, $A$, D, E, G, F, and $N_i$) into the mass formula, masses of tetraquark states for both heavy quarks ($c,b$) have been calculated. We have taken care of the $\xi$ parameter in the mass formula, which acts as the correction factor to the scale parameter $M_0$, which takes different values for different hadronic systems. \\
Firstly, a discussion about singly heavy tetraquarks is carried out by considering their associated quantum numbers. We have predicted their masses, shown in table 4 for the charm quark and table 5 for the bottom quark. We have applied the symmetry that $m_u = m_d$, and thus interchanging the up quark with the down quark does not change the masses of the tetraquarks. Mass prediction for singly heavy tetraquarks provides us a useful comparison with the two newly observed tetraquark candidates $T_{c\bar{s}0}^a (2900)^{++}$ and $T_{c\bar{s}0}(2900)^0$. According to our prediction, they form an isospin triplet having $J^P$ value $1^+$. The predicted mass is 2875.43 $\pm$ 19.00, nearly the same as the experimental mass 2908 $\pm$ 11 $\pm$ 20 MeV. Tables 4 and 5 consists of all the singly heavy tetraquark masses, which can be useful in predicting the tetraquark states.\\
In tables 6 and 7, mass prediction for hidden-charm and hidden-bottom tetraquarks is carried out by considering all the possibilities of quantum number assignments. Also, A comparison with the ref.\cite{63} is carried out for the doubly heavy tetraquarks for both charm and bottom quarks, and we found that for $J^P$ equals to $1^+$, our prediction is 4230.03 $\pm$ 36.80 and in ref.\cite{63}, it is found to be 4190.30 $\pm$ 10.22 for the hidden-charm tetraquarks. A comparison of hidden-charm and hidden bottom tetraquarks is also carried out with the references \cite{29} and \cite{61}. In ref.\cite{29}, masses of hidden-charm tetraquarks have been calculated by considering them as members of the octet, and their comparison with our prediction is shown in table 11. They are also in very close agreement for respective quark contents.\\
Similarly, table 8 is for the masses of triply and fully heavy charm tetraquarks, and table 9 is for the masses of triply and fully heavy bottom tetraquarks. We have compared our prediction with ref.\cite{64} for triply, and fully heavy tetraquarks and masses are in reasonable agreement with the masses predicted in ref.\cite{64}. Also, X(6900) state was experimentally identified by the LHCb in 2020, having an experimental mass of $6886 \pm 11 \pm 11$ MeV, and our calculated mass for this state is 6875.23 $\pm$ 72.95, which is again in excellent agreement with the experimental data. Thus, our analysis of tetraquarks using the extension of the Gursey-Radicati mass formula successfully predicts tetraquark masses. \\
Also, we have plotted the relationship between the evaluated masses and the parameter that are involved into the mass formula such as isospin, hypercharge, and $J^P$ values. Figure 3(a) depicts the behavior of obtained masses with isospin (I), which is similar to that of a parabolic curve in two dimensions, which provides the straightforward hint that the increase in isospin values, increases the mass of the hadrons. Figure 3(b) illustrates the change of masses with spin(S) values, which again shows the parabolic behavior. Figure 3(c) represents the variation of obtained masses with hypercharge (Y) of the tetraquark states, which demonstrates that for negative hypercharge values, masses are more significant, and for positive hypercharge values, mass values are lowered. In section 4, we analyzed the decays of hidden-charm tetraquarks to D-mesons in the framework of the diquark-antidiquark approach. We have listed those decays along with their quantum numbers in table 3. Decay width analysis consists of four hidden-charm states having quark content $u\bar{d}c\bar{c}$, $u\bar{u}c\bar{c}$, $u\bar{s}c\bar{c}$ and $s\bar{d}c\bar{c}$, decays to D mesons by using the formulae listed in ref.\cite{29}. Therefore, our analysis is useful in predicting the masses, $J^P$ values, and decay widths for tetraquark states which are in excellent agreement with the available experimental and theoretical data. \\
\begin{table}
\centering
\begin{tabular}{|c|c|c|c|c|}
    \hline
  S & I & Y 
 & Predicted mass for $N_c$ \\
  \hline
  0 & 1/2 & 1 & 2623.12 $\pm$ 18.72 \\
  \hline
  1 & 1/2 & 1 & 2669.12 $\pm$ 18.88 \\
  \hline
  2 & 1/2 & 1 & 2761.12 $\pm$ 20.06 \\
  \hline
  0 & 0 & 2  & 2416.83 $\pm$ 18.89 \\ 
  \hline
  0 & 1 & 2  & 2480.83 $\pm$ 18.89  \\
  \hline
  1 & 0 & 2  & 2462.83 $\pm$ 19.05  \\
  \hline
  1 & 1 & 2  & 2526.83 $\pm$ 19.05 \\
  \hline
  2 & 0 & 2  &  2554.83 $\pm$ 20.22  \\
  \hline
  2 & 1 & 2  &  2618.83 $\pm$ 20.22  \\
  \hline
  0 & 0 & 0  &  2765.43 $\pm$ 18.67  \\
  \hline
  0 & 1 & 0  &  2829.43 $\pm$ 18.85  \\
  \hline
  1 & 0 & 0  &  2811.43 $\pm$ 18.82  \\
  \hline
  1 & 1  & 0  &  2875.43 $\pm$ 19.00  \\
  \hline
  2 & 0  & 0  &  2903.43 $\pm$ 20.01  \\
  \hline
  2 & 1  & 0  &  2967.43 $\pm$ 20.18 \\
  \hline
  0 & 1/2 & -1 &  2939.73 $\pm$ 18.72 \\
  \hline
   1 & 1/2 & -1 &  2985.72 $\pm$ 18.88 \\
  \hline
   2 & 1/2 & -1 & 3077.72 $\pm$ 20.06 \\
  \hline
   0 & 0 & -2 & 3050.02 $\pm$ 18.89 \\
  \hline
    0 & 1 & -2 & 3114.02 $\pm$ 18.89 \\
  \hline
    1 & 0 & -2 & 3096.02 $\pm$ 19.05 \\
  \hline
    1 & 1 & -2 & 3160.02 $\pm$ 19.05 \\
  \hline
    2 & 0 & -2 & 3188.02 $\pm$ 20.22 \\
  \hline
   2 & 1 & -2 & 3252.02 $\pm$ 20.22 \\
  \hline
  0 & 1/2 & -3 & 3192.33 $\pm$ 19.17 \\
  \hline
  1 & 1/2 & -3 & 3238.33 $\pm$ 19.32  \\
  \hline
  2 & 1/2 & -3 & 3330.33 $\pm$ 20.48  \\
\hline
   \end{tabular}
\caption{Table for $N_c$ = 1 for tetraquark masses in MeV}
\end{table}
\begin{table}
\centering
       \begin{tabular}{|c|c|c|c|c|}
    \hline
  S & I & Y & Predicted mass for $N_b$ \\
  \hline
  0 & 1/2 & -1 & 6405.13 $\pm$ 18.72 \\
  \hline
  1 & 1/2 & -1 & 6451.13 $\pm$ 18.88 \\
  \hline
  2 & 1/2 & -1 & 6543.13 $\pm$ 20.06 \\
  \hline
  0 & 0 & 0  & 6230.83 $\pm$ 18.67 \\ 
  \hline
  0 & 1 & 0  & 6294.83 $\pm$ 18.85  \\
  \hline
  1 & 0 & 0  & 6276.83 $\pm$ 18.82  \\
  \hline
  1 & 1 & 0  & 6340.83 $\pm$ 19.00 \\
  \hline
  2 & 0 & 0  &  6368.83 $\pm$ 20.01  \\
  \hline
  2 & 1 & 0 &  6432.83 $\pm$ 20.18  \\
  \hline
  0 & 0 & -2 &  6515.43 $\pm$ 18.89  \\
  \hline
  0 & 1 & -2  &  6579.43 $\pm$ 18.89  \\
  \hline
  1 & 0 & -2  &  6561.43 $\pm$ 19.05 \\
  \hline
  1 & 1  & -2  &  6625.43 $\pm$ 19.05  \\
  \hline
  2 & 0  & -2  &  6653.43 $\pm$ 20.22  \\
  \hline
  2 & 1  & -2  &  6717.43 $\pm$ 20.22 \\
  \hline
  0 & 1/2 & 1 &  6088.53 $\pm$ 18.72 \\
  \hline
   1 & 1/2 & 1 &  6134.53 $\pm$ 18.88 \\
  \hline
   2 & 1/2 & 1 &  6226.53 $\pm$ 20.06 \\
  \hline
   0 & 0 & 2 &  5882.23 $\pm$ 18.89 \\
  \hline
    0 & 1 & 2 &  5946.23 $\pm$ 18.89 \\
  \hline
    1 & 0 & 2 &  5928.23 $\pm$ 19.05 \\
  \hline
    1 & 1 & 2 &  5992.23 $\pm$ 19.05 \\
  \hline
    2 & 0 & 2 &  6020.23 $\pm$ 20.22 \\
  \hline
   2 & 1 & 2 &  6084.23 $\pm$ 20.22 \\
  \hline
   \end{tabular}
   \caption{Table for $N_b$ = 1 for tetraquark masses in MeV}
\end{table}
  
   \begin{table}[ht]
       \centering
       \begin{tabular}{|c|c|c|c|}
    \hline
  S &  I &  Y & Predicted mass for $N_c$  \\
  \hline
  0 & 0 & 0 & 4120.03 $\pm$ 36.63  \\
  \hline
  0 & 0 & -2 & 4404.63 $\pm$ 36.75 \\
  \hline
  0 & 1  & 0  & 4184.03 $\pm$ 36.73 \\
  \hline 
  1 & 0  & 0  & 4166.03 $\pm$ 36.71 \\
  \hline
  1 & 0  & -2  & 4450.63 $\pm$ 36.83 \\
  \hline
  1 & 1 & 0  & 4230.03 $\pm$ 36.80 \\
  \hline 
  2 & 0 & 0  & 4258.03 $\pm$ 37.33  \\
  \hline
  2 & 0 & -2 & 4542.63 $\pm$ 37.45 \\
  \hline
  2 & 1 & 0 & 4322.03 $\pm$ 37.43 \\
  \hline
  0 & 1/2 & 1 & 3977.73 $\pm$ 36.66  \\
  \hline
  1 & 1/2 & 1 & 4023.73 $\pm$ 36.74  \\
  \hline
  0 & 1/2 & -1 & 4294.33 $\pm$ 36.66  \\
  \hline
  1 & 1/2 & -1 & 4340.33 $\pm$ 36.75 \\
  \hline
  \end{tabular}
  \caption{Table for $N_c$ = 2 for tetraquark masses in MeV}
  \end{table}
  
   \begin{table}[ht]
       \centering
       \begin{tabular}{|c|c|c|c|}
    \hline
  S & I & Y & Predicted masses for $N_b$  \\
  \hline
  0 & 0  & 0 & 11050.83 $\pm$ 109.67 \\
  \hline
  0 & 0 & -2  & 11335.43 $\pm$ 109.71 \\
  \hline
  0 & 1  & 0  & 11114.83 $\pm$ 109.71\\
  \hline 
  1 & 0 & 0  & 11096.83 $\pm$ 109.70 \\
  \hline
  1 & 0 & -2  & 11381.43 $\pm$ 109.74 \\
  \hline
  1 & 1 & 0  & 11160.83 $\pm$ 109.73 \\
  \hline 
  2 & 0 & 0  & 11189.83 $\pm$ 109.91 \\
  \hline
  2 & 0 & -2 & 11473.43 $\pm$ 109.95 \\
  \hline
  2 & 1 & 0 & 11252.83 $\pm$ 109.94 \\
  \hline
  0 & 1/2 & 1 & 10908.53 $\pm$ 109.68 \\
  \hline
  1 & 1/2 & 1 & 10954.53 $\pm$ 109.71 \\
  \hline
  0 & 1/2 & -1 & 11271.73 $\pm$ 109.72 \\
  \hline
  1 & 1/2 & -1 & 11252.83 $\pm$ 109.94 \\
  \hline
  \end{tabular}
  \caption{Table for $N_b$ = 2 for tetraquark masses in MeV}
  \end{table}
  \begin{table}[ht]
      \centering
     \begin{tabular}{|c|c|c|c|}
    \hline
 S & I & Y & Predicted mass for $N_c$ \\
 \hline
  0 & 1/2 & 0 & 5498.63$\pm$ 54.76 \\
  \hline
  1 & 1/2 & 0 & 5544.63 $\pm$ 54.82 \\
  \hline
  2 & 1/2 & 0 & 5636.63 $\pm$ 55.23 \\
  \hline
  0 & 0 & 1 & 5308.33 $\pm$ 54.77 \\
  \hline
  1 & 0 & 1 & 5354.33$\pm$ 54.82 \\
  \hline
  2 & 0 & 1 & 5446.33$\pm$ 55.24 \\
  \hline
  0 & 0 & -1 & 5624.93 $\pm$ 54.77 \\
  \hline
  1 & 0 & -1 & 5670.93 $\pm$ 54.82 \\
  \hline
  2 & 0 & -1  & 5762.93 $\pm$ 55.24  \\
  \hline
  0 & 0 & 0 & 6829.23 $\pm$ 72.91 \\
  \hline
  1 & 0 & 0 & 6875.23$\pm$ 72.95  \\
  \hline
  2 & 0  & 0  & 6967.23 $\pm$ 73.27 \\
  \hline
  \end{tabular}
\caption{Table for $N_c$ = 3 and 4 for tetraquark masses in MeV}
   \end{table}
   \begin{table}[ht]
      \centering
     \begin{tabular}{|c|c|c|c|}
    \hline
 S & I & Y & Predicted mass for $N_b$ \\
 \hline
  0 & 1/2 & 1 & 15728.53 $\pm$ 164.45 \\
  \hline
  1 & 1/2 & 1 & 15774.53 $\pm$ 164.47 \\
  \hline
  2 & 1/2 & 1 & 15866.53 $\pm$ 164.61 \\
  \hline
  0 & 1/2 & -1 & 16045.13 $\pm$ 164.45 \\
  \hline
  1 & 1/2 & -1 & 16091.13 $\pm$ 164.47 \\
  \hline
  2 & 1/2 & -1 & 16183.13 $\pm$ 164.61 \\
  \hline
  0 & 0 & 0 & 15704.53 $\pm$ 164.45 \\
  \hline
  1 & 0 & 0 & 15750.53 $\pm$ 164.67 \\
  \hline
  2 & 0 & 0 & 15842.53 $\pm$ 164.61 \\
  \hline
  0 & 0 & 0 & 20691.83 $\pm$ 219.23\\
  \hline
  1 & 0 & 0 & 20736.83 $\pm$ 219.25 \\
  \hline
  2 & 0 & 0 & 20828.83 $\pm$ 219.35\\
  \hline
  \end{tabular}
       \caption{Table for $N_b$ = 3 and 4 for tetraquark masses in MeV}
   \end{table}
   \begin{table*}[ht] 
   \centering
\begin{tabular}{|c|c|c|c|c|c|}
\hline
S & I & Y & Predicted Masses (MeV) & Ref.\cite{29} prediction & State \\
\hline
0 & 1 & 0 & 4184.03$\pm$ 36.73 & 3.83 GeV & $Z_c^+/Z_c^-/Z_c^0$ \\
\hline
1 & 1 & 0 & 4230.03$\pm$ 36.80 & 3.88 GeV & ,, \\
\hline
2 & 1 & 0 & 4322.03$\pm$ 37.43 & 3.94 GeV &,, \\
\hline
0 & 1/2  & 1  & 3977.73 $\pm$ 36.66 & 4.00 GeV & $Z_{cs}^+/Z_{cs}^0$ \\
\hline
1 & 1/2 & 1 & 4023.73 $\pm$ 36.74 & 4.03 GeV & ,, \\
\hline
2 & 1/2 & 1 & 4115.73 $\pm$ 37.36 & 4.17 GeV & ,, \\
\hline
0 & 1/2 & -1 & 4294.33 $\pm$ 36.66 & 4.04 GeV & $Z_{cs}^-/$ $\bar{Z_{cs}^0}$ \\
\hline
1 & 1/2 & -1 & 4340.33 $\pm$ 36.74 & 4.09 GeV & ,, \\
\hline
2 & 1/2 & -1 & 4432.33$\pm$ 37.36 & 4.17 GeV &,, \\
\hline
0 & 0  & 0 & 4120.03$\pm$ 36.63 & 4.12 GeV & X \\
\hline
1 & 0 & 0 & 4166.03$\pm$ 36.71 & 4.16 GeV & ,, \\
\hline
2 & 0  & 0 & 4258.03 $\pm$ 37.33 & 4.24 GeV & ,, \\
\hline
 \end{tabular} 
\caption{Masses for Tetraquark Octet \cite{29} }
\end{table*}
\begin{figure}[]
\centering
\subfloat[] { \includegraphics[scale=0.6]{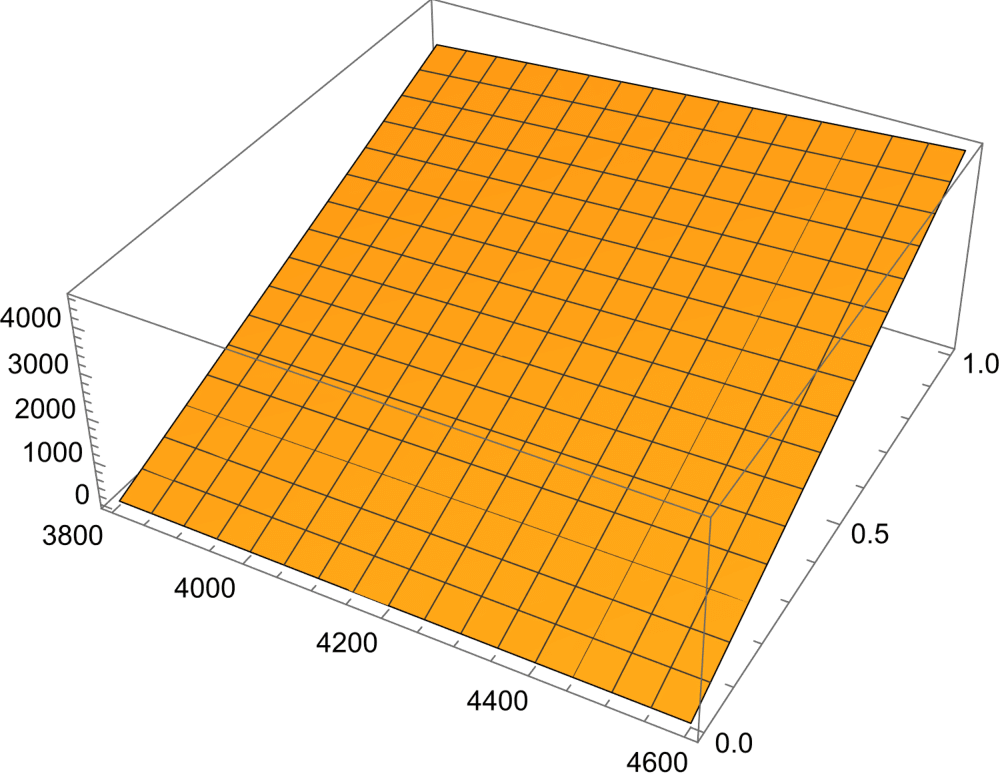} }
\\
\subfloat[] { \includegraphics[scale=0.6]{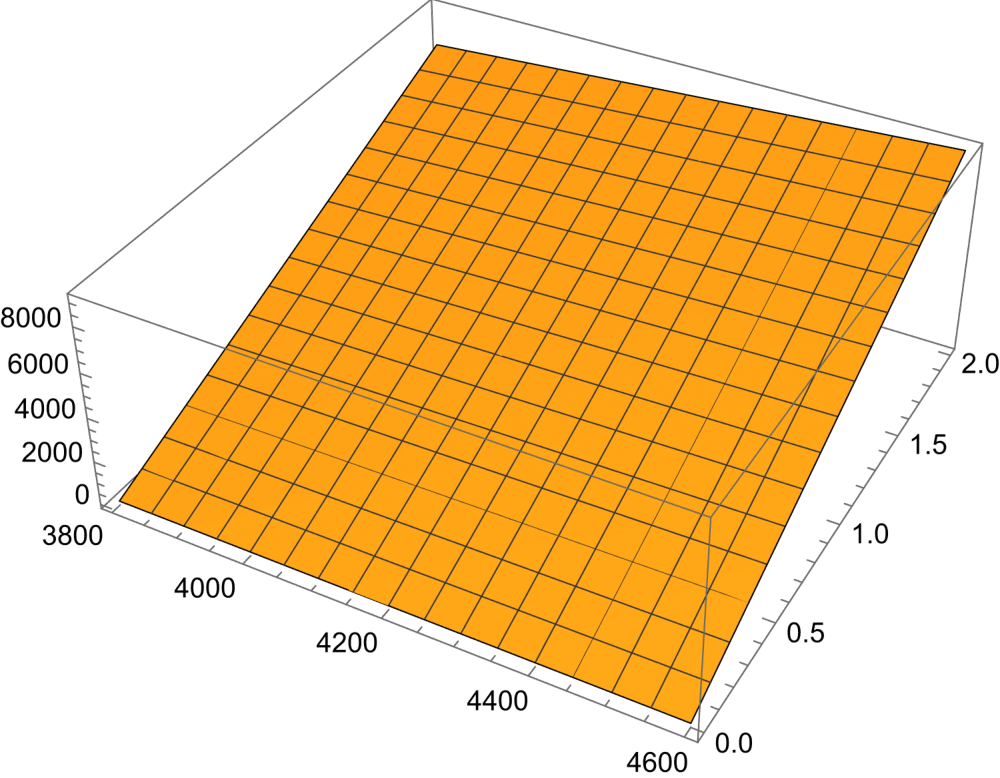} }
\\
\subfloat[] { \includegraphics[scale=0.6]{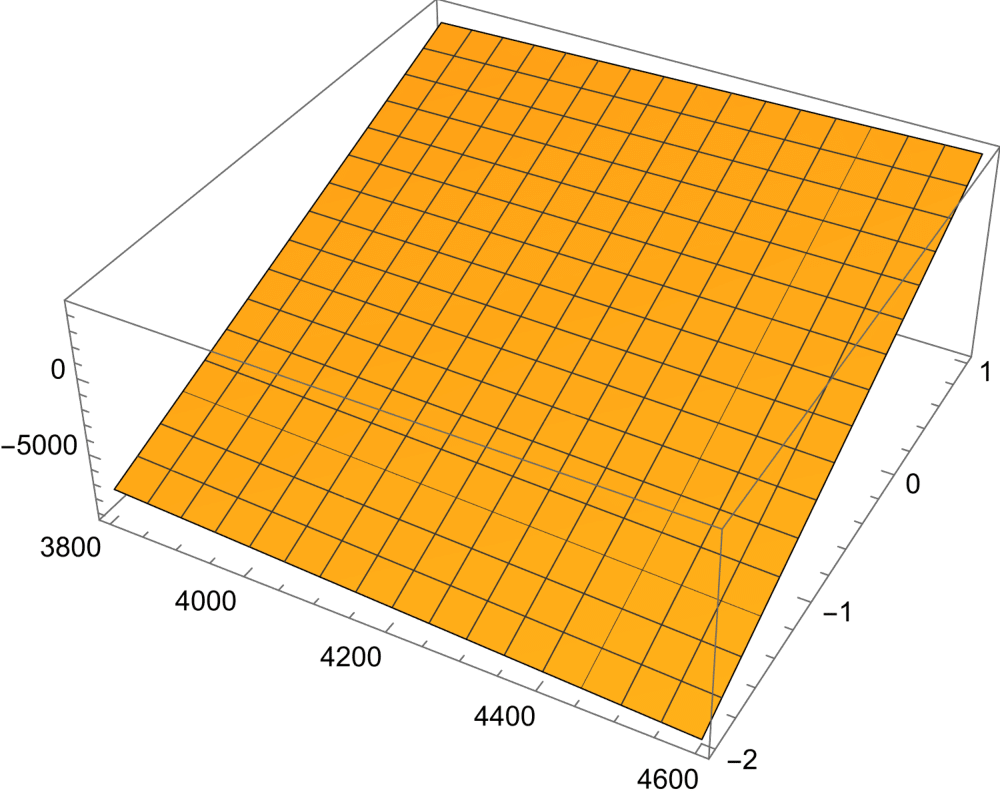} }
\caption{figures 3(a), 3(b), and 3(c) depicts the behavior of the obtained masses with the isospin, spin, and hypercharge, respectively}
\label{fig.3}
\end{figure}
\section{Conclusions}
The LHCb collaboration has recently reported the observation of two singly heavy exotic meson structures having quark content  $u\bar{d}c\bar{s}$ with a significance of 6.5 $\sigma$ and $\bar{u}dc\bar{s}$ with a significance of 8 $\sigma$ were observed in the $D_s^+\pi^+$ and $D_s^+\pi^-$ invariant mass spectra in two B-decay processes $B^+ \rightarrow D^-D_s^+\pi^+$ and $B^0 \rightarrow \bar{D^0}D_s^+ \pi^-$, respectively. The measured mass and width are $M_{exp} = 2908\pm 11 \pm 20$ MeV and $\Gamma_{exp}= 136 \pm 23 \pm 11 $ MeV. The $T_{c\bar{s}0}^a (2900)^{++}$. Our analysis found that the predicted mass is 2875.43 $\pm$ 19.00, nearly the same as the experimental mass for these states. Many works regarding this discovery have been proposed to study these exotic states, such as NRPQM ref.\cite{1} etc. In this present study, we focused on finding the masses of these exotic meson states using the extension of the original GR mass formula, which correctly describes the baryon and pentaquark spectrum and is also able to predict the masses of two newly observed singly heavy tetraquarks by considering them as the member of isospin triplet with $J^P$ value equals to $1^+$. We have classified tetraquarks using $SU(3)_f$ and $SU(2)_s$ symmetries and then with the $SU(6)_{sf}$ representation using the young tableau technique. We considered the SU(3) flavor octet for the tetraquarks and predicted their masses using their quantum numbers. Through our analysis, we have predicted their masses, spin-parity assignments, etc. We have computed the masses of fully heavy charm tetraquarks $(cc\bar{c}\bar{c})$, which is in very close agreement with experimental X(6900) state masses \cite{56} and other theoretical approaches like flip-flop model, butterfly model, etc.\cite{58}. Also, the decay width evaluation for hidden-charm tetraquark states is carried out in terms of the effective coupling $F_{T_{cc}}^2$ by considering the decay of a few of the predicted hidden-charm tetraquark states into D-mesons into the framework of the diquark-antidiquark model.

\section{Acknowledgement}
The author is grateful for the financial support provided by the Department Of Science and Technology (DST), New Delhi, India.\\
Project No. CRG/2019/005159

\end{document}